\documentclass[aps,nobibnotes,two column,amssymb,secnumarabic,superscriptaddress,reprint]{revtex4-1}

\usepackage{amsmath}
\usepackage{amssymb}
\usepackage{bbm}
\usepackage{xcolor}
\usepackage{graphicx}
\usepackage{dcolumn}
\usepackage{bm}
\usepackage{hyperref}

\newcommand{\ket}[1]{\vert #1 \rangle}
\newcommand{\bra}[1]{\langle #1 \vert}

\begin{document}
\title{One-photon Solutions to Multiqubit Multimode quantum Rabi model}
\author{Jie Peng}
\email{jpeng@xtu.edu.cn}
\affiliation{Hunan Key Laboratory for Micro-Nano Energy Materials and Devices and School of
Physics and Optoelectronics, Xiangtan University, Hunan 411105, China}
\author{Juncong Zheng}
\affiliation{Hunan Key Laboratory for Micro-Nano Energy Materials and Devices and School of
Physics and Optoelectronics, Xiangtan University, Hunan 411105, China}
\author{Jing Yu}
\affiliation{International Center of Quantum Artificial Intelligence for Science and Technology (QuArtist) \\
and Physics Department, Shanghai University, 200444 Shanghai, China}
\author{Pinghua Tang}
\affiliation{Hunan Key Laboratory for Micro-Nano Energy Materials and Devices and School of
Physics and Optoelectronics, Xiangtan University, Hunan 411105, China}
\author{G. Alvarado Barrios}
\affiliation{International Center of Quantum Artificial Intelligence for Science and Technology (QuArtist) \\
and Physics Department, Shanghai University, 200444 Shanghai, China}
\author{Jianxin Zhong}
\affiliation{Hunan Key Laboratory for Micro-Nano Energy Materials and Devices and School of
Physics and Optoelectronics, Xiangtan University, Hunan 411105, China}
\author{Enrique Solano}
\email{enr.solano@gmail.com}
\affiliation{International Center of Quantum Artificial Intelligence for Science and Technology (QuArtist) \\
and Physics Department, Shanghai University, 200444 Shanghai, China}
\affiliation{
Department of Physical Chemistry, University of the Basque Country UPV/EHU, Apartado 644, 48080 Bilbao, Spain}
\affiliation{
IKERBASQUE, Basque Foundation for Science, Plaza Euskadi 5, 48009, Spain}
\affiliation{
IQM, Nymphenburgerstr. 86, 80636 Munich, Germany}
\author{F. Albarr\'{a}n-Arriagada}
\affiliation{International Center of Quantum Artificial Intelligence for Science and Technology (QuArtist) \\
and Physics Department, Shanghai University, 200444 Shanghai, China}
\author{Lucas Lamata}
\email{llamata@us.es}
\affiliation{Departamento de F\'isica At\'omica, Molecular y Nuclear, Universidad de Sevilla, 41080 Sevilla, Spain}

\begin{abstract}
General solutions to the quantum Rabi model involve subspaces with unbounded number of photons. However, for the multiqubit multimode case, we find special solutions with at most one photon for arbitrary number of qubits and photon modes. Unlike the Juddian solution, ours exists for arbitrary single qubit-photon coupling strength with constant eigenenergy. This corresponds to a horizontal line in the spectrum, while still being a qubit-photon entangled state. As a possible application, we propose an adiabatic scheme for the fast generation of arbitrary single-photon multimode W states with nonadiabatic error less than $1\%$. Finally, we propose a superconducting circuit design, showing the experimental feasibility of the multimode multiqubit Rabi model.

\end{abstract}
\maketitle
\emph{Introduction.--}
The quantum Rabi model \cite{rabi,jc} describes the interaction between a two-level system and a single photonic mode at the most fundamental level. Despite its simple form, the exact solution of the model was not found until 2011~\cite{braak}. Since the quantum Rabi model involves both rotating and counter-rotating interaction terms, all Fock states are connected and there is no closed subspace, turning the Hamiltonian hard to solve.

The quantum Rabi model plays an important role in quantum optics~\cite{verlang,jorge,jsp,lucas}, molecular physics~\cite{molecular}, condensed matter physics~\cite{irish} and quantum information~\cite{blais,xyl}. However, in most applications we need to consider more than one qubit and/or more than one mode. For instance, to perform a controlled gate~\cite{romero}, essential for  universal quantum computing~\cite{ab,deng}, while multimode models are useful for the generation of the multipartite entangled states~\cite{sr,jos}. Therefore, a mathematical description with physical implications for models with more than one qubit and one mode is essential for the development of scalable and efficient protocols, suitable for current technology demands.

In this article, we find special solutions with at most one photon to the multiqubit multimode quantum Rabi model (MMQRM) for arbitrary number of qubits and modes, although the interaction terms still connect all photon number states. Unlike the well known Juddian solution \cite{judd} of the quantum Rabi model, these quasi-exact solutions exist for arbitrary single qubit-photon coupling strength with constant eigenenergy, corresponding to a horizontal line in the spectrum, while still being a qubit-photon entangled state. This coherent superposition is what makes the photon population trapped in zero and one, and we call it a special dark state~\cite{scully}. Furthermore, we use such solutions to propose a fast entangled state generation protocol, where we simultaneously obtain a two-qubit Bell state and an arbitrary single-photon $M-$mode W state $
  \vert W \rangle_{M}=\sum_{i=1}^M g_i\vert 0_1 0_2\cdots 1_i 0_{i+1}\cdots 0_M \rangle
$~\cite{zl,ybs,asaa} through adiabatic passage. Here, $g_{i}$ just corresponds to the coupling strength between the qubits and the $i$-th photon mode. It is known that W states are robust under
particle loss \cite{mm} and a central resource in several quantum information processing protocols \cite{pa,sbz,ll,eak,lhy}. Consequently, various schemes have been presented to generate them \cite{me, men,bf,bozou,gpg,zjd,gcg,yhk,vm}. Due to the reach of ultrastrong coupling and peculiarities of the special dark states, the most interesting advantage of using the MMQRM is the fast generation (less than 70 $\omega^{-1}$)  and high fidelity (larger than $99\%$), outperforming the fastest two-qubit CPHASE gate (30-45~ns) to date~\cite{rol} for $\omega=3$ GHz. The generation time will not change with the number of modes $M$ and no external laser is needed. Since $g_i$ is adjustable, we can generate any W state in a unified and convenient way, such as the perfect W states which are useful in quantum teleportation \cite{pa}. Finally, we propose a superconducting circuit design to show the experimental feasibility of catch and release of these W states. This result paves the way to the implementation of fast protocols in quantum information using the MMQRM.

\emph{Special quasi-exact solutions to the multiqubit multimode quantum Rabi model.--}\label{s3}
We present our method to obtain the quasi-exact solution with at most one photon
of the multiqubit and multimode quantum Rabi model
\begin{equation}\label{mode}
H_{pq}=\sum_{i=1}^M\omega_i a_i^\dagger a_i+\sum_{i=1}^M\sum_{j=1}^N g_{ij} \sigma_{jx}(a_i+a^\dagger_i)+\sum_{j=1}^N \Delta_j \sigma_{jz},
\end{equation}
where $a^{\dagger}_i$ and $a_i$ are the $i$-th photon mode creation and annihilation operators with frequency $\omega_i$, respectively. Also, $\sigma_{j \alpha} (\alpha=x,y,z)$ are the Pauli matrices corresponding to the $j$-th qubit, $2\Delta_j$ is the energy level
splitting of the $j$-th qubit, and $g_{ij}$ is the qubit-photon coupling parameter between the $i$-th mode and $j$-th qubit. 

Since Hamiltonian~(\ref{mode}) breaks the $U(1)$ symmetry, there is no closed subspace consisting of finite photon number states. However, it has a $\mathbb{Z}_2$ symmetry with generator $R=\exp[i\pi\sum_{i=1}^M
a^\dagger_i a_i]\Pi_j \sigma_{jz}$.
Accordingly, we categorize all N-qubit states
$\{|\psi_{Nq}\rangle\}$ into two sets corresponding to the eigenvuales of
$\Pi_j \sigma_{jz}$, being $1$ and $-1$, and denote them  $2^{N-1}$ dimensional row vectors, $|\psi_{Nq+}\rangle$ and $|\psi_{Nq-}\rangle$, respectively. We also denote all k-photon states with M modes by $|k_M\rangle$. Hence, there are two invariant subspaces
\begin{eqnarray}
|0_M, \psi_{Nq+}\rangle\leftrightarrow |1_M,
\psi_{Nq-}\rangle\leftrightarrow |2_M,
\psi_{Nq+}\rangle\leftrightarrow \cdots \\
|0_M, \psi_{Nq-}\rangle\leftrightarrow |1_M,
\psi_{Nq+}\rangle\leftrightarrow |2_M,
\psi_{Nq-}\rangle\leftrightarrow \cdots 
\end{eqnarray}
with positive and negative parity, respectively.
So $H_{pq}$ will take the following form in the $\pm$ parity subspace 
 \begin{eqnarray}\label{off}
H_{pq}^\pm=\left(
  \begin{array}{cccccc}
   D_{0}^\pm &O_{0}^\pm &0 & 0&0&\dots\\
    O_{0}^\pm & D_{1}^\pm&O_{1}^\pm & 0 &0&\dots  \\
   0&O_{1}^\pm &D_{2}^\pm & O_{2}^\pm &0&\dots \\
     \dots&\dots&\dots&\dots&\dots&\dots\\
  \end{array}
\right),
\end{eqnarray}
where $D_{k}^\pm$ and $O_{k}^\pm$ ($k=0,1,2,3,\ldots$) are ${2^{N-1}C_{M+k-1}^{k}\times 2^{N-1}C_{M+k-1}^{k}}$
matrices \cite{sl}. 
Although all Fock states are connected by the interaction terms, it is possible to find some special solutions with finite photon number. Indeed, there are special dark states in the single-mode multiqubit Rabi model consisting of finite Fock states~\cite{pj,pj1,pj2}. 

We search now for solutions with at most $L$ photons taking the form $|\psi_\pm\rangle=c^\pm_{0M}|0_M,\psi_{Nq,\pm}\rangle+c^\pm_{1M}|1_M,\psi_{Nq,\pm}\rangle+\ldots+c^\pm_{LM}|L_M,\psi_{Nq,\pm}\rangle$ in the MMQRM by solving the energy eigenequation 
\begin{equation}\label{pqe}
(H_{pq}^\pm-E^\pm)|\psi_\pm\rangle=0.
\end{equation} 
Substituting $H_{pq}^\pm$ from Eq. \eqref{off} into Eq. \eqref{pqe} and solving for the coefficients  $c^\pm_{k_M}$,  we see there are more equations than variables ($c^\pm_{k_M}$) provided we do not truncate $H_{pq}$. Nevertheless, solutions
could exist when the Hamiltonian parameters themselves meet certain condition to satisfy Eq.~\eqref{pqe}. Indeed, we have proved that for $L>1$ the condition for a solution takes the general form $f(g_{ij},\omega_i,\Delta_j)=0$ \cite{sl}, just like the Juddian solution for the single-qubit quantum Rabi model. However, we find a remarkable case $L=1$, where the condition reduces to $f(\omega_i,\Delta_j)=0$ and $f(g_{ij})=0$, so the solutions change from single points into horizontal lines in the spectra when $\omega_i$ and $\Delta_j$ are fixed. The latter may have important applications in quantum
\begin{figure}[htbp]
\center
\resizebox{0.9\columnwidth}{!}{
  \includegraphics{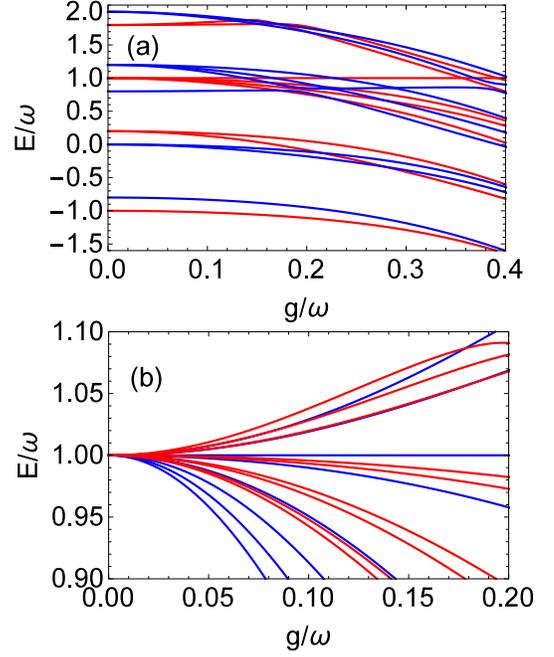}
}
\renewcommand\figurename{\textbf{Figure}}
\caption[2]{(a) Spectrum of the two-qubit two-mode quantum Rabi model with $\omega_1=\omega_2=\omega$, $\Delta_1=0.9\omega$, $\Delta_2=0.1\omega$, $g_{11}=g_{12}=g_{21}=g_{22}=g$. (b) Spectrum of the three-qubit two-mode quantum Rabi model with $\Delta_1=\Delta_2=\Delta_3=\omega_1=\omega_2=\omega$, $g_{11}=g_{12}+g_{13}=g_{21}$, $g_{12}=g_{22}=g_{13}=g_{23}$. Red lines correspond to even parity while blue lines to odd parity. 
\label{fig1}}
\end{figure} information, as will be discussed below. Therefore, we will focus on the case $L=1$, where Eq. \eqref{pqe} reduces to
\begin{eqnarray}\label{eig1}
\left(
  \begin{array}{cc}
   D_{0}^\pm-E^\pm &0 \\
    O_{~0}^\pm & D_{~1}^\pm-E^\pm  \\
      0& O_{~1} \\
  \end{array}
\right)\left(
  \begin{array}{c}
   c^\pm_{0_M} \\
   c^\pm_{1_M} \\
  \end{array}
\right)=0 
\end{eqnarray}
after elementary row matrix transformation. For the single qubit and single mode case, $O_1$ is just a c-number, so there is no nontrivial solution for $O_1^\pm c^\pm_{11}=0$. However, for the multiqubit and multimode case, this equation can be satisfied if matrix $O_1^\pm$ has eigenvalue $0$ with $c^\pm_{1M}$ being its corresponding eigenvector. Then, we make the elementary row transformation to the matrix in Eq.~\eqref{eig1}, and the solution is obtained if there are more columns than nonzero rows. At the same time, the condition for the parameters is calculated.

For the two-qubit and M-mode case, we find the special solution for even parity to be (see \cite{sl} for details)
\begin{equation}\label{dk1}
|\psi_{2+}\rangle=\frac{1}{{\cal
N}}\left[(\Delta_1-\Delta_2)|0_M,\uparrow,\uparrow\rangle+|W_M\rangle(|\downarrow,\uparrow\rangle-
 |\uparrow,\downarrow\rangle)
\right]
\end{equation}
{where $|W_M\rangle= g_1|1,0,0,\ldots,0\rangle+g_2|0,1,0,\ldots,0\rangle+\dots+g_M|0,0,0,\ldots,1\rangle$ with the condition $\omega_i=\omega$ for all $i$, $g_{ij}=g_i$ for all $j$ and $\Delta_1+\Delta_2=\omega=E^+$. This special solution has some novel properties: (1) It exists for arbitrary $g_i$, with constant eigenenergy $E^+=\omega$, corresponding to a horizontal line in the spectrum, as shown in Fig. \ref{fig1} (a).  (2) It is a special dark state because although the interaction terms connect all Fock states, the population is trapped in vacuum and single photon multimode W states~\cite{scully}. There are two other similar dark-state solutions corresponding to odd parity which are described in the supplementary material~\cite{sl}.

For the three-qubit and M-mode case, the quasi-exact solution for odd parity reads
\begin{eqnarray}\label{2}
|\psi_{3-}\rangle=&&|W_M\rangle(|\uparrow,\downarrow,\downarrow\rangle-|\downarrow,\uparrow,\downarrow\rangle
-|\downarrow,\downarrow,\uparrow\rangle+|\uparrow,\uparrow,\uparrow\rangle)\nonumber\\
&&+\frac{\omega g_{13}}{g_{12}}|0_M,\uparrow,\uparrow,\downarrow\rangle+\frac{\omega g_{12}}{g_{13}}|0_M,\uparrow,\downarrow,\uparrow\rangle\nonumber\\
&&-\frac{\omega g_{11}^2}{g_{12}g_{13}}|0_M,\downarrow,\uparrow,\uparrow\rangle ,
\end{eqnarray}
where $|W_M\rangle= g_{11}|1,0,0,\ldots,0\rangle+g_{21}|0,1,0,\ldots,0\rangle+\dots+g_{M1}|0,0,0,\ldots,1\rangle$ with  the conditions
$\Delta_j=\omega_i=\omega=E^-, g_{i1} = g_{i2}+g_{i3}$. This eigenstate corresponds to the horizontal line $E/\omega=1$ in Fig. \ref{fig1} (b).

As the qubit number grows, the existence condition will be harder to satisfy,
but there are indeed quasi-exact solutions for the N-qubit M-mode Rabi model formed by the product of the two-qubit singlet Bell state $|\psi_B\rangle=\frac{1}{\sqrt{2}}(|\downarrow\uparrow\rangle-|\uparrow\downarrow\rangle)$ and $|\psi_2\rangle$ or $|\psi_3\rangle$,
\begin{eqnarray}
|\psi_N\rangle=|\psi_2\rangle\otimes(|\psi_{B}\rangle)^{(N-2)/2}, \label{Nq1} \\
|\psi_N\rangle=|\psi_3\rangle\otimes(|\psi_{B}\rangle)^{(N-3)/2} ,
\end{eqnarray}
for even and odd $N$, respectively.

\emph{Fast Generation of the arbitrary single-photon multimode W state.--}
Here, the special dark state solution for the two-qubit and M-mode quantum Rabi model $|\psi_{2+}\rangle$ in Eq. \eqref{dk1} is an excellent candidate for generating arbitrary W states $|W_M\rangle$ through adiabatic passage: (1) The energy gap limiting the adiabatic speed can be much larger than normal cases due to its peculiarities, which will be detailed later. (2) It consists of only $|0_M\rangle$ and $|W_M\rangle$ for the photon part, and the coefficient of the former is proportional to $\Delta_1-\Delta_2$, so once the qubit frequencies are tuned to be equal, we immediately arrive at $|W_M\rangle$ for any nonzero $g_i$, robust against its fluctuation. 
(3)~The coefficient of $|0_10_2\cdots 1_i0_{i+1}\cdots 0\rangle$ is just $g_i$ in $|W_M\rangle$, so it is very convenient to construct any W state with arbitrary modes by adjusting $g_i$ directly, and the generation time will stay the same. (4) At the beginning of the adiabatic passage, the initial state $|0_M,\uparrow,\uparrow\rangle$ is easy to prepare, while in the end, we arrive at a W state $|W_M\rangle$ and a qubit Bell state $|\psi_B\rangle$ simultaneously, both very useful in quantum information processing, while no external laser pulse is needed. Thereafter, the W state is automatically stored in the resonator because the photon field and qubit Bell state are decoupled. 
\begin{figure}[htbp]
\center
\resizebox{1\columnwidth}{!}{
  \includegraphics{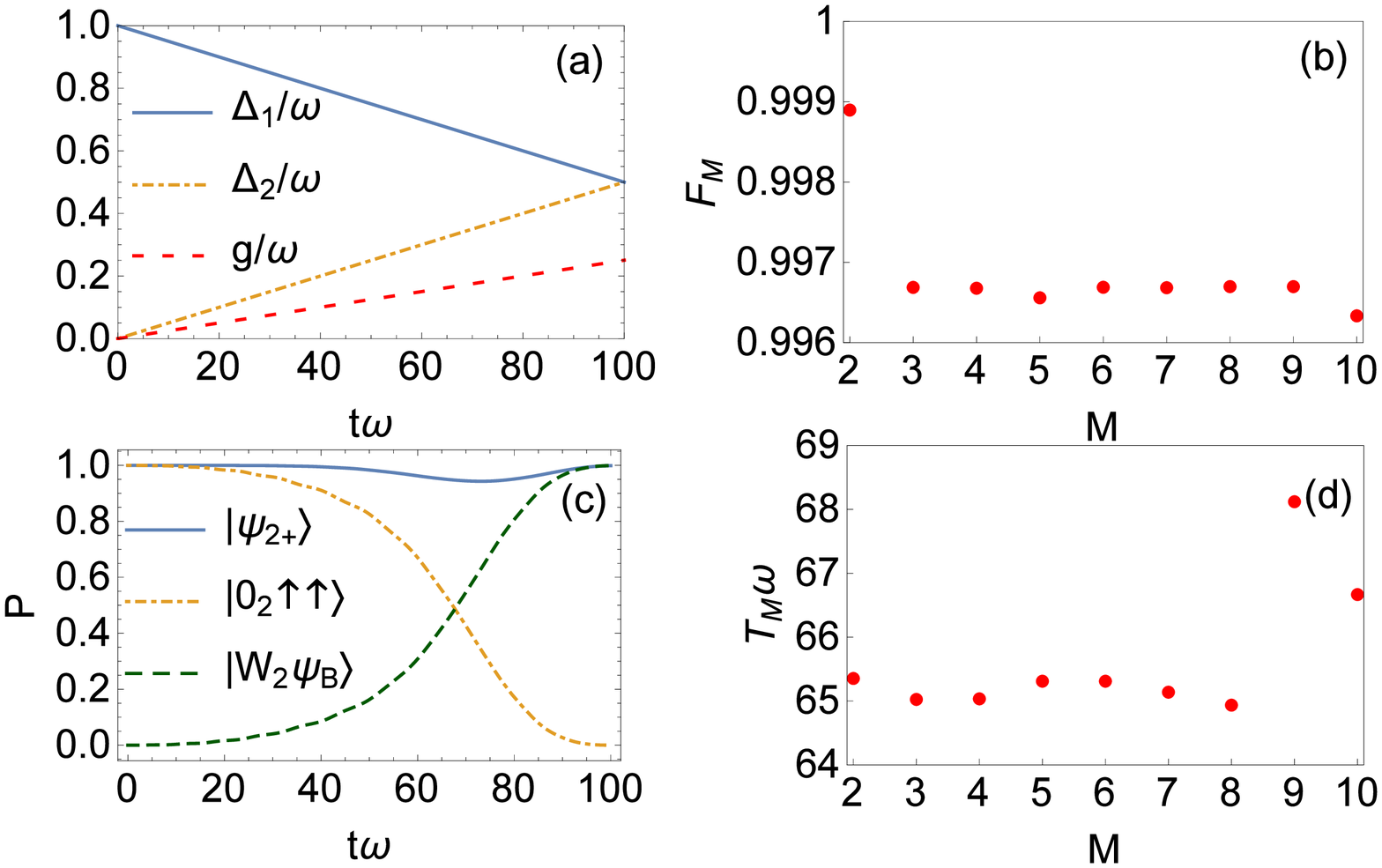}
}
\renewcommand\figurename{\textbf{Figure}}
\caption[2]{Adiabatic evolution of the special dark state $| \psi_{2+} \rangle$ of Eq.~\eqref{dk1} from $|0_M\uparrow\uparrow\rangle$ to $|W_M\psi_B\rangle$ for the two-qubit M-mode quantum Rabi model, where $|\psi_B\rangle=\frac{1}{\sqrt{2}}(|\downarrow\uparrow-\uparrow\downarrow\rangle)$, and $|W_M\rangle$ is the prototype W state with all $g_i$'s equal. (a) Adiabatic trajectory used
to vary the parameters for $M=2$. $g_1=g_2=g$.  (b) Fidelity $F_M=|\langle \psi(T)|W_M\rangle|^2$ when $T$ is fixed to 100$\omega^{-1}$ for mode number $M$. (c) Population of different states during the adiabatic process for $M=2$. (d) Interaction time $T_M$ to reach $F_M>0.99$ for mode $M$ in unit of $\omega^{-1}$.
\label{fig2}}
\end{figure}

Our scheme is as follows. First, two qubits are excited by pumping pulses and coupled to M resonators in vacuum states with initial coupling strength $g_{i1} = g_{i2} = g_i =0$. The qubit frequencies are non-identical and always satisfy
$\Delta_1 + \Delta_2 = \omega_i = \omega$. Then, we slowly decrease $|\Delta_1-\Delta_2|$ to $0$ while increase $g_i$ to a nonzero value, so that the target state $|W_M\psi_B\rangle$ is obtained. As an example, the numerical simulation for the adiabatic evolution of the two-qubit two-mode case is shown in Fig. \ref{fig2} (c) with the adiabatic trajectory used
to vary the parameters shown in Fig. \ref{fig2} (a). The evolution time is just $T=100\omega^{-1}$ and the fidelity $F_2=|\langle\psi(T)|W_2\psi_B\rangle|^2$ reaches $99.89\%$. If the evolution time is fixed to $100\omega^{-1}$, the fidelities $F_M$ for the M-mode case are shown in Fig. \ref{fig2} (b), which are almost equal and higher than $99.6\%$. If the fidelities are restricted to be higher than $99\%$, the time costs for each mode are shown in Fig. \ref{fig2} (d), which are less than $69\omega^{-1}$. 
 
According to the current available circuit QED technology, the transmon frequency $\Delta/\pi$ can be tuned from 0 to 6 GHz \cite{mdh,pk}, hence the resonator frequency $\omega/2\pi$ is chosen to be 3 GHz \cite{blais1} to satisfy $\Delta_1\pm\Delta_2=\omega$ at $g/2\pi=0$. The maximum value of $g/2\pi$ can  be tuned to 0.7535 GHz \cite{blais1}, that is, $0.2505\omega/2\pi$. Therefore, our scheme shown in Fig. \ref{fig2} (a) is within experimental reach and the adiabatic evolution takes only 33.3 ns with nonadiabatic error $0.1\%$. It is faster than the current state-of-the-art two-qubit CPHASE gate with operation time of 40~ns~\cite{Barends,mk}, and comparable to the fastest two-qubit gates to date of 30-45~ns~\cite{rol}. If we restrict the nonadiabatic error to less than $1\%$, then the average evolution time to generate $|W_M\psi_B\rangle$ from $|0_M \uparrow \uparrow\rangle$ for $M=2,3,\ldots,10$ will be 21.9 ns. Note that we have chosen the simplest linear adiabatic path shown in Fig.~\ref{fig2} (a), but we can also consider a ``faster adiabatic'' trajectory to reduce the time such as in Refs. \cite{Barends,jmm,yuchen}.
 
\begin{figure}[htbp]
\center
\resizebox{1\columnwidth}{!}{
  \includegraphics{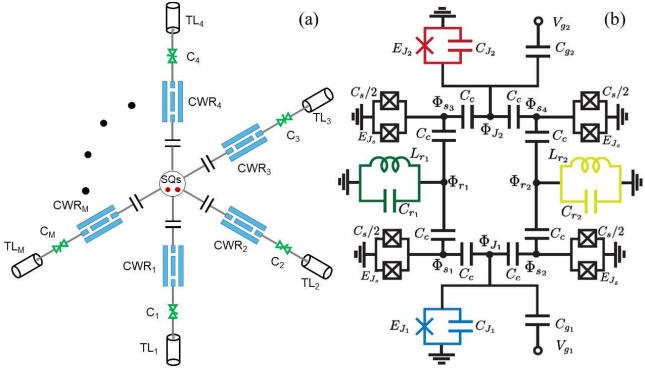}
}
\renewcommand\figurename{\textbf{Figure}}
\caption[2]{(a) Schematic setup for the generation and release of the W state: Two SQs are capacitively coupled to M CWRs. Each CWR is connected to a TL through a variable coupler C, such that the photon emission rate into the TL is controllable. (b) A superconducting circuit design for the two-qubit two-mode Rabi model.
\label{fig3}}
\end{figure}

Although we did not optimize the adiabatic path, the simulated adiabatic evolution speed had already been faster than or similar to the optimized ones~\cite{Barends,jmm,yuchen}. This is due to the reach of ultrastrong coupling and the peculiarities of the special dark state $|\psi_{2+} \rangle$ \cite{sl}. (1) There are $C_{M+1}^2+1$ degenerate eigenstates $\vert\psi_{E=\omega}\rangle$ including $\vert\psi_{2+}\rangle$ at Jaynes-Cummings coupling regime where the rotating wave approximation is applied. (2) $\langle\psi_{E=\omega}|\dot{H}|\psi_{2+}\rangle=0$, no matter how fast the parameters change. As can be seen in the spectrum in Fig. \ref{fig1} (a), there are three energy levels very close to the horizontal line dark state $|\psi_{2+}\rangle$ at 
$E=\omega$ when $g$ is small. Actually, they correspond to degenerate eigenstates of $H_{pq}$ with $E=\omega$ when rotating wave approximation is applied \cite{sl}, which is valid for small $g$, and we proved $\langle\psi_{E=\omega}|\dot{H}|\psi_{2+}\rangle=0$ \cite{sl}, such that the actual energy gap limiting the adiabatic speed according to the adiabatic theorem ~\cite{fest,born}
\begin{equation}
\left|\frac{\langle E_{m}(t)|\dot{H}|E_{n}(t)\rangle}{(E_m-E_n)^2}\right|\ll1 , ~~~
m\neq n , ~~~ t\in[0,T].
\end{equation}  is about $|\Delta_1-\Delta_2|$ at Jaynes-Cummings coupling regime, which can be tuned to be $\omega$, much larger than normal cases. For the same reason, the adiabatic speed is not limited by the vanishing energy gap at the degeneracy points around  
 $g\approx0.357$ and $0.392$. Although the W state and two-qubit Bell state are obtained once $\Delta_1=\Delta_2$ and $g\neq0$, we find the adiabatic evolution will be faster if $g$ is increased to the ultrastrong coupling regime. 

\begin{figure}[htbp]
\center
\resizebox{1\columnwidth}{!}{
  \includegraphics{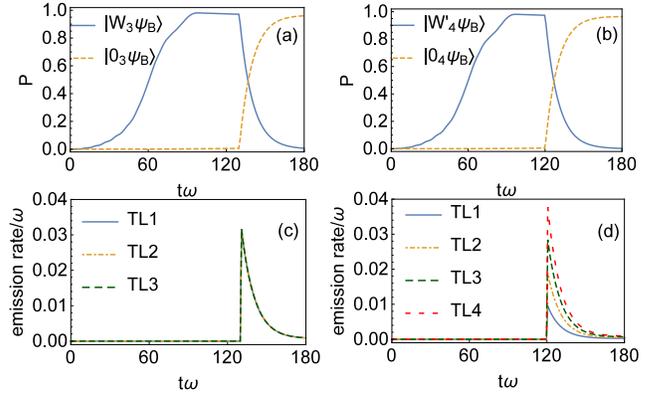}
}
\renewcommand\figurename{\textbf{Figure}}
\caption[2]{Numerical simulation for the catch and release of the prototype three-mode W state $|W_3\rangle=\frac{1}{\sqrt{3}}(|100\rangle+|010\rangle+|001\rangle)$ (left panel) and the four-mode W state $|W^\prime_4\rangle=\frac{1}{\sqrt{10}}(|1000\rangle+\sqrt{2}|0100\rangle+\sqrt{3}|0010\rangle+2|0001\rangle)$ (right panel). (a) and (b) Population of different states inside resonators. (c) and (d)~Emission rates into transmission lines.
\label{fig4}}
\end{figure}

\emph{Catch and release of the W state.}--
On the other hand, just generating a W state inside resonators is not convenient for its transport and detection, such that we propose a scheme to store or extract them out on demand for practical usage in quantum teleportation and related tasks. Our scheme is depicted in Fig. \ref{fig3} (a), where two superconducting qubits (SQs) in the center of the devices are capacitively coupled to N coplanar waveguide resonators (CWRs), which can be described by the two-qubit M-mode Rabi model. A detailed description of the superconducting circuit design for the two-qubit two-mode Rabi model (see Fig. \ref{fig3} (b)) is shown in \cite{sl}. Besides, there is an externally variable coupler to modulate the decay rate $\kappa_c$ of each CWR through that coupler, such that its photon emission into the connected transmission line (TL) is controllable. In current experimental setups, $\kappa_c$ can be tuned to be 1000 times the CWR intrinsic decay rate $\kappa_{in}$ \cite{yin}, such that we can catch (generate and store) and release the W state on demand. 

At the first step, an arbitrary single photon M-mode W state and a two-qubit Bell state can be generated simultaneously using the scheme discussed above. Thereafter, the qubit Bell state is decoupled from the CWRs, hence $|W_M\rangle$ is naturally stored in resonators, and robust against any fluctuation on $g_i$. After a desired time $\tau$, we turn on the coupling $\kappa_c$ and therefore the W state is released into the transmission lines. For realistic experimental considerations, we choose the intrinsic  dissipation rate for each resonator $\kappa_{in}=10^{-4}\omega$ and  $\kappa_{c}=10^{-1}\omega$. The energy relaxation rate for each qubit $\gamma_j=10^{-5}\omega$ and the dephasing rate $\gamma_{j\phi}=10^{-4}\omega$. Using a Lindblad equation \cite{sl}, we present the numerical simulation in Fig. \ref{fig4}. The time cost for generating $|W_3\rangle$ and $|W'_4\rangle$ is still $100\omega^{-1}$, but the fidelity is reduced to $98\%$ due to dissipation, damping and dephasing. 
\begin{figure}[htbp]
\center
\resizebox{0.9\columnwidth}{!}{
  \includegraphics{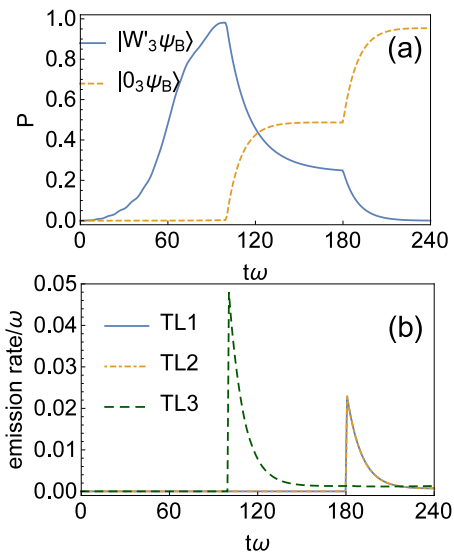}
}
\renewcommand\figurename{\textbf{Figure}}
\caption[2]{Numerical simulation for the generation and controlled release of the perfect three-mode W state $|W^\prime_3\rangle=\frac{1}{2}(|100\rangle+|010\rangle+\sqrt{2}|001\rangle)$ for quantum teleportation. (a) Population of different states inside resonators. (b) Emission rates into transmission lines. 
\label{fig5}}
\end{figure}

However, the prototype W state $|W_3\rangle=|100\rangle+|010\rangle+|001\rangle$ generated in Fig. \ref{fig4} (c) cannot be used to perform quantum teleportation, while the so-called perfect W state $|W^\prime_3\rangle=\frac{1}{2}(|100\rangle+|010\rangle+\sqrt{2}|001\rangle)$ can \cite{pa}. This state can be easily generated just by tuning $g_3=\sqrt{2}g_2=\sqrt{2}g_1$, as shown in Fig. \ref{fig5} (a). The first two qubits are kept by Alice while the third one is sent to Bob to carry out the remote communication. If we know the distance $L$ between them, we can delay the emission of the W state into the first two transmission lines by $L/C$ to assure they receive the qubits at the same time, which will increase the communication security. The corresponding numerical simulation is shown in Fig. \ref{fig5} (b).

\emph{Conclusions.--} All Fock states are excited by the dipole interaction in the ultrastrong coupling regime of the quantum Rabi model, where the rotating-wave approximation is not valid. Therefore, although the operation can be faster, it seems impossible to construct any kind of single photon state.  However, we find that for the multiqubit multimode quantum Rabi model, there exist special dark eigenstates consisting of only vacuum and single photon multimode W states for the photon part in the whole coupling regime with constant energy. Accordingly, we propose a unified scheme to adiabatically generate arbitrary W states using the special dark-state solution to the two-qubit M-mode quantum Rabi model, being able to take advantages of the ultrastrong coupling and avoid its dynamical complexities. Due to the peculiarities of these states, the energy gap limiting the speed can be tuned to be much larger than normal cases. Hence, the time cost according to the current circuit QED technology (33ns with nonadiabatic error $0.1\%$ and 21.9 ns with nonadiabatic error $1\%$) is comparable to the fastest two-qubit gate (30-45ns) to date \cite{rol}. Optimization of the adiabatic path could further accelerate the adiabatic process in our protocol. The coefficient of $|0_1,0_2,\ldots,1_i,\dots\rangle$ in the W state is just proportional to the coupling strength between the qubits and the $i$-th resonator, such that any W state can be obtained just by tuning $g_i$. Moreover, it can also be released into the transmission lines on demand, which are illustrated to be useful in quantum information processing. Similar uses of other special dark states still need to be explored.

\emph{Acknowledgements.--}
This work was supported by the National Natural Science Foundation of China (11704320), Natural Science Foundation of Hunan Province, China ( 2018JJ3482), the National
Basic Research Program of China (2015CB921103), the Program for Changjiang Scholars and Innovative Research Team in University (No. IRT13093), the funding from PGC2018-095113-B-I00, PID2019-104002GB-C21, Spanish Government PID2019-104002GB-C22 (MCIU/AEI/FEDER, UE), Spanish Government PGC2018-095113-B-I00 (MCIU/AEI/FEDER, UE), Basque Government IT986-16, as well as from QMiCS (820505) and OpenSuperQ (820363) of the EU Flagship on Quantum Technologies, EU FET Open Grant Quromorphic (828826), EPIQUS (899368) and Shang- hai STCSM (Grant No. 2019SHZDZX01-ZX04).

\newpage
\begin{widetext}

\subsection*{}
{\bf \large Supplementary Material: One-photon Solutions to Multiqubit Multimode quantum Rabi model}

\renewcommand{\thesection}{S\arabic{section}}
\renewcommand{\thesubsection}{\Alph{subsection}}
\renewcommand{\thesubsubsection}{\alph\arabic{subsubsection}}
\renewcommand{\theequation}{S\arabic{equation}}
\renewcommand{\thefigure}{S\arabic{figure}}
\renewcommand{\thetable}{S\arabic{table}}
\setcounter{equation}{0}
\setcounter{figure}{0}

~\\

This supplementary material contains four parts: (1) Quasi-exact solution to the multiqubit multimode Rabi model; (2) Peculiarities of the special dark state $|\psi_{2+} \rangle$; (3) Demonstration of the circuit design for the implementation of the two-qubit two-mode quantum Rabi model with variable couplings; (4) The Lindblad master equation we used for numerical simulation.

\section{Quasi-exact solution to the multiqubit multimode quantum Rabi model}
$D_{k}^\pm$ and $O_{k}^\pm$ ($k=0,1,2,3,\ldots$) can be
written as
\begin{eqnarray}\tiny
D_{k}^\pm &=&(\langle
k_M,\psi_{Nq,\pm(-1)^k}|)^T H_{pq}(|k_M,\psi_{Nq,\pm(-1)^k}\rangle),\label{d}\\
O_{k}^\pm &=&(\langle (k+1)_M,\psi_{Nq,\mp(-1)^k}|)^T
H_{pq}(|k_M,\psi_{Nq,\pm(-1)^k}\rangle),\label{o}
\end{eqnarray}
where $(|k_M,\psi_{Nq,\pm(-1)^j}\rangle)$ is a $2^{N-1}C_{M+k-1}^{k}$ dimensional row
vector, consisting of all possible qubit-photon product state with $k$ photons , $N$ qubits and $\pm$ parity. Hence $D_{k}^\pm$, $O_{k}^\pm$ are $2^{N-1}C_{M+k-1}^{k}\times 2^{N-1}C_{M+k-1}^{k}$
matrices. 
If eigenstate $|\psi_\pm\rangle=c^\pm_{0M}|0_M,\psi_{Nq,\pm}\rangle+c^\pm_{1M}|1_M,\psi_{Nq,\pm}\rangle+\ldots+c^\pm_{LM}|L_M,\psi_{Nq,\pm}\rangle$ exists, then according to Eq. \eqref{pqe}, we obtain
\begin{eqnarray}\label{eig}
\left(
  \begin{array}{cccccc}
   D_{0}^\pm-E^\pm &O_{0}&0 & 0&\dots\\
    O_{0} & D_{1}^\pm-E^\pm&O_{1}& 0 &\dots  \\
      \dots&\dots&\dots&\dots&\dots\\
      0&\dots&O_{L-2} &D_{L-1}^\pm-E^\pm &O_{L-1}^\pm \\
   0&\dots&0&O_{L-1} &D_{L}^\pm-E^\pm  \\
    0&\dots&0&0 & O_{L}^\pm \\
  \end{array}
\right)\left(
  \begin{array}{c}
   c^\pm_{0,M} \\
   c^\pm_{1,M} \\
      \ldots\\
   c^\pm_{L-1,M}\\
   c^\pm_{L,M} \\
  \end{array}
\right)=0.
\end{eqnarray}
Clearly, there are more equations than variables, but quasi-exact solution is possible if the parameters meet certain condition. A necessary but not sufficient condition for Eq. \eqref{eig} is
\begin{eqnarray}\label{eiga}
\left|
  \begin{array}{cccccc}
   D_{0}^\pm-E^\pm &O_{0}&0 & 0&\dots\\
   O_{0} & D_{1}^\pm-E^\pm&O_{1}& 0 &\dots  \\
      \dots&\dots&\dots&\dots&\dots\\
      0&\dots&O_{L-2} &D_{L-1}^\pm-E^\pm &O_{L-1} \\
   0&\dots&0&O_{L-1} &D_{L}^\pm-E^\pm
  \end{array}
\right|=0 .
\end{eqnarray}
Consequently, in general, $E^\pm$ is dependent on the couplings. But if $L=1$, Eq.~\eqref{eig} reduces to
\begin{eqnarray}\label{1pho}
\left(
  \begin{array}{cc}
   D_{0}^\pm-E^\pm &0 \\
    O_{~0}^\pm & D_{~1}^\pm-E^\pm  \\
      0& O_{~1} \\
  \end{array}
\right)\left(
  \begin{array}{c}
   c^\pm_{0_M} \\
   c^\pm_{1_M} \\
  \end{array}
\right)=0 ,
\end{eqnarray}
after elementary row transformations, which will be demonstrated later, such that the determinant in Eq. \eqref{eiga} reduces to
\begin{eqnarray}\label{eigb}
\left|
  \begin{array}{cccccc}
   D_{0}^\pm-E^\pm &0\\
   O_{0} & D_{1}^\pm-E^\pm 
  \end{array}
\right|=0 .
\end{eqnarray}
Here, $E^\pm$ is independent of the couplings and its existence condition reduces to $f(\omega_i,\Delta_j)=0$ and $f(g_{ij})=0$, which will be much easier to realize than  $f(\omega_i,\Delta_j,g_{ij})=0$ generally. This character may have important applications in quantum information processing, such that we focus on the case $L=1$.

Let us start with the simplest case, the two-qubit two-mode quantum Rabi model,
\begin{equation}\label{2q2m}
H_{2q2m}=\omega_1 a_1^\dagger a_1+\omega_2 a_2^\dagger a_2+ (g_{11} \sigma_{1x}+g_{12} \sigma_{2x})(a_1+a^\dagger_1)+ (g_{21} \sigma_{1x}+g_{22} \sigma_{2x})(a_2+a^\dagger_2)+\Delta_1 \sigma_{1z}+\Delta_2\sigma_{2z}.
\end{equation}
For even parity, in the basis formed by $\{|0,0,\uparrow,\uparrow\rangle,|0,0,\downarrow,\downarrow\rangle,|1,0,\uparrow,\downarrow\rangle,|1,0,\downarrow,\uparrow\rangle,|0,1,\uparrow,\downarrow\rangle,|0,1,\downarrow,\uparrow\rangle,|2,0,\uparrow,\uparrow\rangle,|2,0,\downarrow,\downarrow\rangle,|1,1,\uparrow,\uparrow\rangle,|1,1,\downarrow,\downarrow\rangle,|0,2,\uparrow,\uparrow\rangle,|0,2,\downarrow,\downarrow\rangle\}$, the coefficient matrix of Eq. \eqref{eig} reads
\begin{eqnarray}
\small\label{eigc}\nonumber
  \left(\begin{array}{cccccc}
  -\Delta_1-\Delta_2-E^+ &0&g_{11} & g_{12}&g_{21}&g_{22} \\
   0&\Delta_1+\Delta_2-E^+ &g_{12}&g_{11}& g_{22}&g_{21} \\
   g_{11} & g_{12}& \omega_1+\Delta_1-\Delta_2-E^+ &0&0&0 \\
    g_{12} & g_{11}&0& \omega_1-\Delta_
    1+\Delta_2-E^+&0&0 \\
     g_{21} & g_{22}& 0&0& \omega_2+\Delta_1-\Delta_2-E^+&0 \\
      g_{22} & g_{21}&0&0&0&\omega_2-\Delta_1+\Delta_2-E^+ \\
       0&0&\sqrt{2}g_{11} & \sqrt{2}g_{12}&0&0 \\
         0&0&\sqrt{2}g_{12} & \sqrt{2}g_{11}&0&0 \\
         0&0&g_{21} & g_{22}&g_{11} & g_{12} \\
          0&0&g_{22} & g_{21}&g_{12} & g_{11} \\
           0&0&0 & 0&\sqrt{2}g_{21} & \sqrt{2}g_{22} \\
            0&0&0 & 0&\sqrt{2}g_{22} & \sqrt{2}g_{21} \\
  \end{array}\right) .
\end{eqnarray}
Obviously, it takes the form of Eq. \eqref{1pho} after elementary row transformations. Nontrivial solutions exist if there are less nonzero rows than columns, which will be satisfied when $\omega_1=\omega_2=E^+=\Delta_1+\Delta_2$, $g_{11}=g_{12}=g_1$, and $g_{21}=g_{22}=g_2$. The solution reads
\begin{equation}\label{dka}
|\psi_{22+}\rangle=\frac{1}{{\cal
N}}\left[(\Delta_1-\Delta_2)|0,0,\uparrow,\uparrow\rangle+(g_1|1,0\rangle+g_2|0,1\rangle)(|\downarrow,\uparrow\rangle-
 |\uparrow,\downarrow\rangle)
\right].
\end{equation}

Extending our analysis to the M-mode case, it is easy to find the special dark state solution $|\psi_{2+}\rangle$ \eqref{dk1} in the main text. For odd parity, there are similar special dark states
\begin{eqnarray}
|\psi\rangle_{2-,a}=\frac{1}{{\cal
N^\prime}}\left[(\Delta_1+\Delta_2)|0_M,\uparrow,\downarrow\rangle+|W_M\rangle(
 |\downarrow,\downarrow\rangle
-|\uparrow,\uparrow\rangle)\right],\label{dk2}\\
 |\psi\rangle_{2-,b}=\frac{1}{{\cal
N^\prime}}\left[(\Delta_1+\Delta_2)|0_M,\downarrow,\uparrow\rangle+|W_M\rangle
 |\downarrow,\downarrow\rangle
-|\uparrow,\uparrow\rangle\right]\label{dk3},
\end{eqnarray}
with the condition $\Delta_1-\Delta_2=\omega=E^-$ and $\Delta_2-\Delta_1=\omega=E^-$ respectively, and the condition for $\omega_i, g_{ij}$ is the same as the even parity. Using a similar method, we can obtain the special dark state solution in Eq. \eqref{2} for the three-qubit M-mode quantum Rabi model.

\section{Peculiarities of the special dark state $|\psi_{2+}\rangle$}

There are two peculiarities of $|\psi_{2+}\rangle$: 1. $\langle\psi_{E=\omega}|\dot{H}|\psi_{2+}\rangle=0$, no matter how fast the parameters changes. 2. There are $C_{M+1}^2+1$ degenerate eigenstates $\vert\psi_{E=\omega}\rangle$ at Jaynes-Cummings coupling regime where the rotating wave approximation is applied. Here we give a proof. 

First, we prove $\langle\psi_{E=\omega}|\dot{H}|\psi_{2+}\rangle=0$. In the adiabatic evolution of the two-qubit and two-mode quantum Rabi model with $\Delta_1+\Delta_2=\omega_1=\omega_2=\omega$, $g_{12}=g_{11}=g_1$and $g_{22}=g_{21}=g_2$,
\begin{equation}
 \dot{H}_{2R}=\dot{\Delta}_1\sigma_{1z}-\dot{\Delta}_1\sigma_{2z}+\dot{g}_{1}(a_1+a_1^\dag)\sigma_{1x} + \dot{g}_{1}(a_1+a_1^\dag)\sigma_{2x} + \dot{g}_{2}(a_2+a_2^\dag)\sigma_{1x} + \dot{g}_{2}(a_2+a_2^\dag)\sigma_{2x} .
\end{equation}
So it is easy to find 
\begin{equation}\label{co}
\dot{H}_{2R}|\psi_{2+}\rangle=\frac{1}{{\cal
N}}(|\downarrow,\uparrow\rangle+
 |\uparrow,\downarrow\rangle)\left[(\dot{g}_1(\Delta_1-\Delta_2)-2\dot{\Delta}_1g_1)|1,0\rangle+(\dot{g}_2(\Delta_1-\Delta_2)-2\dot{\Delta}_1g_2)|0,1\rangle
\right].
\end{equation}
Substituting $E=\omega$ into $(H_{2R}-E)|\psi_{E=\omega}\rangle=0$, we have
\begin{eqnarray}
(\Delta_2-\Delta_1)\langle 1,0,\downarrow,\uparrow|\psi_{E=\omega}\rangle=(\Delta_1-\Delta_2)\langle 1,0,\uparrow,\downarrow|\psi_{E=\omega}\rangle\nonumber\\
(\Delta_2-\Delta_1)\langle 0,1,\downarrow,\uparrow|\psi_{E=\omega}\rangle=(\Delta_1-\Delta_2)\langle 0,1,\uparrow,\downarrow|\psi_{E=\omega}\rangle .
\end{eqnarray}
Therefore,
\begin{eqnarray}\label{coe}
 \langle 1,0,\downarrow,\uparrow|\psi_{E=\omega}\rangle=-\langle1,0,\uparrow,\downarrow|\psi_{E=\omega}\rangle,~~~~
 \langle 0,1,\downarrow,\uparrow|\psi_{E=\omega}\rangle=-\langle 0,1,\uparrow,\downarrow|\psi_{E=\omega}\rangle,
\end{eqnarray}
such that $\langle\psi_{E=\omega}|\dot{H}_{2R}|\psi_{2+}\rangle=0$, no matter how fast the parameters change. In consequence, the adiabatic speed is not restricted by the adiabatic theorem
\begin{equation}
\left|\frac{\langle E_{m}(t)|\dot{H}|E_{n}(t)\rangle}{(E_m-E_n)^2}\right|\ll1~~~
m\neq n ~~~t\in[0,T]
\end{equation} at the degeneracy point. This result can be easily extended to the two-qubit and M-mode case.

Next, we prove there are four degenerate eigenstates for the two-qubit two-mode Jaynes-Cummings (JC) model
\begin{equation}\label{2q2m}
H_{2J}=\omega_1 a_1^\dagger a_1+\omega_2 a_2^\dagger a_2+ g_{11}(a_1\sigma_1^++a_1^\dag\sigma_1) + g_{12}(a_1\sigma_2^++a_1^\dag\sigma_2) + g_{21}(a_2\sigma_1^++a_2^\dag\sigma_1) + g_{22}(a_2\sigma_2^++a_2^\dag\sigma_2)+\Delta_1 \sigma_{1z}+\Delta_2\sigma_{2z} ,
\end{equation} 
when $\Delta_1+\Delta_2=\omega_1=\omega_2$, $g_{12}=g_{11}=g_1$ and $g_{22}=g_{21}=g_2$. For this model, the excitation number operator $C= a_1^\dagger a_1+ a_2^\dagger a_2+(\sigma_{1z}+\sigma_{2z})/2+1$ is conserved. In this manner, in the subspace for $C=2$ consisting of $\{|0,0,\uparrow,\uparrow\rangle,|1,0,\uparrow,\downarrow\rangle,|1,0,\downarrow,\uparrow\rangle,|0,1,\uparrow,\downarrow\rangle,|0,1,\downarrow,\uparrow\rangle,|2,0,\downarrow,\downarrow\rangle,|1,1,\downarrow,\downarrow\rangle,|0,2,\downarrow,\downarrow\rangle\}$, the Hamiltonian reads
\begin{eqnarray}\small\label{eigd}
  \left(\begin{array}{cccccccc}
  \omega &g_{1} & g_{1}&g_{2}&g_{2}&0&0&0\\
  g_1&\omega+\Delta_1-\Delta_2 &0&0&0&\sqrt{2}g_{1}&g_{2}& 0\\
   g_{1} &0& \omega-\Delta_1+\Delta_2&0&0&\sqrt{2}g_{1}&g_{2}&0\\
    g_{2} &0&0& \omega+\Delta_1-\Delta_2&0&0&g_{1}&\sqrt{2}g_{2}\\
     g_{2} & 0& 0&0& \omega-\Delta_1+\Delta_2&0&g_1&\sqrt{2}g_2\\
      0 &\sqrt{2} g_{1}&\sqrt{2} g_{1}&0&0&\omega&0&0\\
       0& g_{2} & g_{2}&g_1&g_1&0&\omega &0\\
         0&0&0&\sqrt{2}g_{2} & \sqrt{2}g_{2}&0&0&\omega\\
  \end{array}\right),
\end{eqnarray}
and the secular equation $|H_{2J}-E|=0$ takes the form
\begin{equation}
(E-\omega)^4 f(E,\Delta_1,\Delta_2,g_1,g_2,\omega)=0.
\end{equation}
We have then four degenerate
eigenstates $|\psi_{E=\omega}\rangle$ with $E=\omega$, one of which can be $|\psi_{2+}\rangle$ in Eq. \eqref{dk1}. 
It is easy to find that Eqs. \eqref{co} and \eqref{coe} still hold for $H_{2J}$, so we still obtain $\langle\psi_{E=\omega}|\dot{H}_{2J}|\psi_{2+}\rangle=0$. Therefore, the energy gap limiting the adiabatic speed according to the adiabatic theorem will only be dependent on the other four energy levels, that is, $|\Delta_1-\Delta_2|$, which can be tuned to be $\omega$ when $g\sim 0$. This result can be easily extended to the two-qubit M-mode quantum Rabi model, and that is one reason why the adiabatic generation of the multimode W state is exceptionally fast. Since there are $C_{M+1}^2$ states
$|2_M,\downarrow,\downarrow\rangle$ which has the same energy $\omega$ as $\vert 0_M,
\uparrow,\uparrow\rangle$ when $\Delta_1+\Delta_2=\omega$ and $g_{ij}=g_i$, it is easy to prove there are $C_{M+1}^2+1$ degenerate eigenstates with $E=\omega$.

\section{Demonstration of the circuit design for the implementation of the two-qubit two-mode quantum Rabi model with variable coupling}
For the physical implementation of the two-qubit two-mode quantum Rabi model, we propose the circuit given by Fig. \ref{fig3} (b), which is described by the Lagrangian
\begin{eqnarray}\nonumber
\mathcal{L} = && \sum^2_{j=1}\bigg[\frac{C_{g_j}}{2}(\dot{\Phi}_{J_j}-V_{g_j})^2+\frac{C_{J_j}}{2}\dot{\Phi}_{J_j}^2+E_{J_j}\cos{\varphi_{J_j}}\bigg]+\sum^2_{j=1}\bigg[\frac{C_{r_j}}{2}\dot{\Phi}_{r_j}^2-\frac{\Phi_{r_j}^2}{2L_{r_j}}\bigg]+\sum^4_{j=1}\bigg[\frac{C_{s}}{2}\dot{\Phi}_{s_j}^2+2E_{J_s}\cos{(\varphi^{(j)}_{ext})}\cos{\varphi_{s_j}}\bigg]
\\\nonumber
&&+\frac{C_c}{2}(\dot{\Phi}_{J_1}-\dot{\Phi}_{s_1})^2+\frac{C_c}{2}(\dot{\Phi}_{J_1}-\dot{\Phi}_{s_2})^2
+\frac{C_c}{2}(\dot{\Phi}_{r_1}-\dot{\Phi}_{s_1})^2+\frac{C_c}{2}(\dot{\Phi}_{r_2}-\dot{\Phi}_{s_2})^2+\frac{C_c}{2}(\dot{\Phi}_{J_2}-\dot{\Phi}_{s_3})^2+\frac{C_c}{2}(\dot{\Phi}_{J_2}-\dot{\Phi}_{s_4})^2\\
&&+\frac{C_c}{2}(\dot{\Phi}_{r_1}-\dot{\Phi}_{s_3})^2+\frac{C_c}{2}(\dot{\Phi}_{r_2}-\dot{\Phi}_{s_4})^2,
\label{Eq1}
\end{eqnarray}
where the Josephson phase is $\varphi_j={2\pi\Phi_j}/{\Phi_0}$, with $\Phi_0=h/2e$  the superconducting flux quantum, and $2e$ is the electrical charge of a Cooper pair. 

By applying the Legendre transformation $\mathcal{H}(\Phi_j,Q_j)=\sum_j Q_j\dot{\Phi}_j-\mathcal{L}$ and making some simplifications, we obtain the Hamiltonian 
\begin{eqnarray}\nonumber
\mathcal{H} = && \frac{1}{\bar{C}_{J_1}}(Q_{J_1}-2e\bar{n}_{g_1})^2-E_{J_1}\cos{(\varphi_{J_1})}
+\gamma_{J_1}(\sin{\varphi_{J_1}})^2+\frac{1}{\bar{C}_{J_2}}(Q_{J_2}-2e\bar{n}_{g_2})^2-E_{J_2}\cos{(\varphi_{J_2})}
+\gamma_{J_2}(\sin{\varphi_{J_2}})^2\\\nonumber
&&+\frac{1}{2\bar{C}_{r_1}}Q_{r_1}^2+\frac{\Phi^2_{r_1}}{2L_{r_1}}+\gamma_{r_1}\Phi^2_{r_{1}}+\frac{1}{2\bar{C}_{r_2}}Q_{r_2}^2+\frac{\Phi^2_{r_2}}{2L_{r_2}}+\gamma_{r_2}\Phi^2_{r_{2}}-\gamma_{J_1r_1}\sin{\varphi_{J_1}}\Phi_{r_{1}}-\gamma_{J_1r_2}\sin{\varphi_{J_1}}\Phi_{r_{2}}\\
&&-\gamma_{J_2r_1}\sin{\varphi_{J_2}}\Phi_{r_{1}}-\gamma_{J_2r_2}\sin{\varphi_{J_2}}\Phi_{r_{2}},
\label{Eq17}
\end{eqnarray}
where $Q_j$ is the conjugate momenta (node charge) for each node,
\begin{eqnarray}
\bar{C}_{J_j}=C_j+2C_c,\quad \bar{C}_{r_j}=C_{r_j}+2C_c, \quad \bar{n}_{g_j}=-\frac{C_{g_j}V_{g_j}}{2e},
\end{eqnarray}
and 
\begin{eqnarray}\nonumber
&& \gamma_{J_1}=\bigg[\frac{1}{E_{J_s}\cos{(\varphi^{(1)}_{ext}})}+\frac{1}{E_{J_s}\cos{(\varphi^{(2)}_{ext}})}\bigg]\frac{C^2_{c}E^2_{J_1}}{4(C_{1}+2C_c)^2},\quad \gamma_{J_2}=\bigg[\frac{1}{E_{J_s}\cos{(\varphi^{(3)}_{ext}})}+\frac{1}{E_{J_s}\cos{(\varphi^{(4)}_{ext}})}\bigg]\frac{C^2_{c}E^2_{J_2}}{4(C_{2}+2C_c)^2} ,
\\ \nonumber
&& \gamma_{r_1}=\bigg[\frac{1}{E_{J_s}\cos{(\varphi^{(1)}_{ext}})}+\frac{1}{E_{J_s}\cos{(\varphi^{(3)}_{ext}})}\bigg]\frac{C^2_{c}\Phi^2_0}{16\pi^2(C_{r_{1}}+2C_c)^2L^2_{r_{1}}}, \\ \nonumber
 && \gamma_{r_2}=\bigg[\frac{1}{E_{J_s}\cos{(\varphi^{(2)}_{ext}})}+\frac{1}{E_{J_s}\cos{(\varphi^{(4)}_{ext}})}\bigg]\frac{\Phi^2_0C^2_{c}}{16\pi^2(C_{r_{2}}+2C_c)^2L^2_{r_{2}}},
\\ \nonumber
&& \gamma_{J_1r_1}=\frac{\Phi_0C^2_{c}E_{J_1}}{4\pi (C_{r_{1}}+2C_c)(C_{1}+2C_c)L_{r_{1}}E_{J_s}\cos{(\varphi^{(1)}_{ext}})},\quad
\gamma_{J_1r_2}=\frac{\Phi_0C^2_{c}E_{J_1}}{4\pi(C_{r_{2}}+2C_c)(C_{1}+2C_c)L_{r_{2}} E_{J_s}\cos{\varphi^{(2)}_{ext}}},\\
&& \gamma_{J_2r_1}=\frac{\Phi_0C^2_{c}E_{J_2}}{4\pi (C_{r_{1}}+2C_c)(C_{1}+2C_c)E_{J_s}\cos{\varphi^{(3)}_{ext}}L_{r_{1}}},\quad
\gamma_{J_2r_2}=\frac{\Phi_0C^2_{c}E_{J_2}}{4\pi(C_{r_{2}}+2C_c)(C_{1}+2C_c) E_{J_s}\cos{\varphi^{(4)}_{ext}}L_{r_{2}}} \, .
\end{eqnarray}

Then, we quantize this Hamiltonian by promoting the classical variables to quantum operators,  $Q_j\rightarrow \hat{Q}_j=2e\hat{n}_j$ and $\varphi_j\rightarrow \hat{\varphi}_j$ with the commutation relation $[e^{i\hat{\varphi}_{j}},\hat{n}_{j}]=e^{i\hat{\varphi}_{j}}$, and $Q_{r_j}\rightarrow \hat{Q}_{r_j}=\sqrt{\hbar \omega_{r_j}\bar{C}_{r_j}/2}(a^\dagger+a)$, and $\Phi_{r_j}\rightarrow \hat{\Phi}_{r_j}=i\sqrt{\hbar \omega_{r_j}\bar{L}_{r_j}/2}(a-a^\dagger)$ with the commutation $[a,a^\dagger]=1$, where $\bar{L}_{r_j}=\frac{1}{1/L_r+2\gamma_{r_j}}$, and $\omega_{r_j}=\sqrt{1/\bar{C}_{r_j}\bar{L}_{r+j}}$. 

Thus, the quantum Hamiltonian reads
\begin{eqnarray}
\hat{\mathcal{H}}&=&\sum_{j=1}^2\hat{\mathcal{H}}_{q}^j+\sum_{j=1}^2\hat{\mathcal{H}}_{r}^j-\gamma_{J_1r_1}\sin{\varphi_{J_1}}\Phi_{r_{1}}-\gamma_{J_1r_2}\sin{\varphi_{J_1}}\Phi_{r_{2}}
-\gamma_{J_2r_1}\sin{\varphi_{J_2}}\Phi_{r_{1}}-\gamma_{J_2r_2}\sin{\varphi_{J_2}}\Phi_{r_{2}} .
\label{EqA20}
\end{eqnarray}
Here,
\begin{eqnarray}
\hat{\mathcal{H}}_{q}^j&=&4E_{C_j}(\hat{n}_{j}-\bar{n}_{g_j})^2-E_{J_j}\cos{(\hat{\varphi}_j)}+\gamma_{J_j}(\varphi_{ext})\sin{(\hat{\varphi}_j)}^2 
\label{EqA21}
\end{eqnarray}
is the Hamiltonian of the $j$th qubit with charge energy $E_{C_j}={e^2}/{2\bar{C}_{J_i}}$, and $\hat{\mathcal{H}}_{r}^j=\hbar \omega_{r_j}a_j^\dagger a_j$ is the Hamiltonian of the $j$th resonator.
Moreover, in the charge number basis, it means $\hat{n}_j=\sum_m m \ket{m_j}\bra{m_j}$, where $|m_j\rangle$ is the $m$th excited state of the $j$th subsystem, $\cos{(\hat{\varphi}_j)}$ and $\sin{(\hat{\varphi}_j)}$ read
\begin{eqnarray}
\cos{(\hat{\varphi}_j)}=\frac{1}{2}\left(\sum_{m}|m_j\rangle\langle m_j+1| + {\rm H.c.} \right),\quad \sin{(\hat{\varphi}_j)}=-\frac{i}{2}\left(\sum_{m}|m_j\rangle \langle m_j + 1 | - {\rm H.c.} \right).
\label{EqA22}
\end{eqnarray}

In the following discussion, we consider $\bar{n}_{g_1}=\bar{n}_{g_2}=0.5$ and $\hbar=1$. Note that the free Hamiltonian of the subsystem $\mathcal{H}^j_{q}$ in Eq.~(\ref{EqA21}) includes both the pure CPB free Hamiltonian and the nonlinear term proportional to $\sin{(\hat{\varphi}_{J_j})}^2$ term. However, as long as we keep $E_{J_j}/E_{C_j}$ in the charge regime, the increase of $\gamma_{J_j}$ will not destroy the anharmonicity of our system. The level of anharmonicity still depends on the ratio $E_{J_j} /E_{C_j}$. Thus, in the charge regime, we can safely perform the two-level approximation. Therefore, the operator $\sin{\hat{\varphi}_{J_j}}$ in the subsystem basis now reads
\begin{eqnarray}
\sin{\hat{\varphi}_{J_j}}=\frac{1}{2}\sigma^y_j ,
\end{eqnarray}
where $\sigma_j^\alpha$ is the Pauli matrix and $\mathbb{I}_j$ is the identity operator. Accordingly, the nonlinear term $\gamma_{J_j}\sin{(\hat{\varphi}_{J_j})}^2$, as displayed in Eq.~(\ref{EqA21}), can be approximated as
\begin{eqnarray}
\gamma_{J_j}\sin{(\hat{\varphi}_{J_j})}^2\approx\frac{\gamma_{J_j}}{4}\mathbb{I},
\end{eqnarray}
which only provides a shift to the qubit frequency. Finally, we obtain the simplified Hamiltonian as follows
\begin{eqnarray}\nonumber
\hat{\mathcal{H}}&=&\frac{\omega_{q_1}}{2}\sigma_1^z+\frac{\omega_{q_2}}{2}\sigma_2^z+\omega_{r_1}a_1^\dagger a_1+\omega_{r_2}a_2^\dagger a_2-i\tilde{\gamma}_{J_1r_1}\sigma_1^y(a_1-a_1^\dagger)-i\tilde{\gamma}_{J_1r_2}\sigma_1^y(a_2-a_2^\dagger)\\
&-&i\tilde{\gamma}_{J_2r_1}\sigma_2^y(a_1-a_1^\dagger)-i\tilde{\gamma}_{J_2r_2}\sigma_2^y(a_2-a_2^\dagger) ,
\label{EqA27}
\end{eqnarray}
where $\omega_{q_1}=E_{J_1}$ and $\omega_{q_2}=E_{J_2}$, and the effective coupling strength
\begin{eqnarray}\nonumber
\tilde{\gamma}_{J_1r_1}&=&\frac{\Phi_0C^2_{c}E_{J_1}}{8\pi (C_{r_{1}}+2C_c)(C_{1}+2C_c)L_{r_{1}}E_{J_s}\cos{(\varphi^{(1)}_{ext}})}\sqrt{\frac{\hbar \omega_{r_1}\bar{L}_{r_1}}{2}},\nonumber\\
\tilde{\gamma}_{J_1r_2}&=&\frac{\Phi_0C^2_{c}E_{J_1}}{8\pi(C_{r_{2}}+2C_c)(C_{1}+2C_c)L_{r_{2}} E_{J_s}\cos{\varphi^{(2)}_{ext}}}\sqrt{\frac{\hbar \omega_{r_2}\bar{L}_{r_2}}{2}},\nonumber\\
\tilde{\gamma}_{J_2r_1}&=&\frac{\Phi_0C^2_{c}E_{J_2}}{8\pi (C_{r_{1}}+2C_c)(C_{1}+2C_c)E_{J_s}\cos{\varphi^{(3)}_{ext}}L_{r_{1}}}\sqrt{\frac{\hbar \omega_{r_1}\bar{L}_{r_1}}{2}},\nonumber\\
\tilde{\gamma}_{J_2r_2}&=&\frac{\Phi_0C^2_{c}E_{J_2}}{8\pi(C_{r_{2}}+2C_c)(C_{1}+2C_c) E_{J_s}\cos{\varphi^{(4)}_{ext}}L_{r_{2}}}\sqrt{\frac{\hbar \omega_{r_2}\bar{L}_{r_2}}{2}}.
\end{eqnarray}
Now, we consider the external flux $\varphi^{(j)}_{ext}$ to be composed by a DC signal and a small AC signal as $\varphi^{(j)}_{ext}=\varphi^{(j)}_{ext}(t)=\varphi^{(j)}_{DC}+\varphi^{(j)}_{AC}(t)$, where 
\begin{eqnarray}
\varphi^{(j)}_{AC}(t)=A^{(j)}_1\cos{(\nu^{(j)}_1 t +\tilde{\varphi}^{(j)}_1)}+A^{(j)}_2\cos{(\nu^{(j)}_2 t +\tilde{\varphi}^{(j)}_2)},
\label{EqA28}
\end{eqnarray}
with $|A^{(j)}_{1}|, |A^{(j)}_{2}|\ll |\varphi_{DC}|$. Then, we can approximate
\begin{eqnarray}
\frac{1}{E_{J_s}\cos{(\varphi^{(j)}_{ext})}}\approx\frac{1}{\bar{E}_{J_s}}\left[1+\frac{\sin{({\varphi}^{(j)}_{DC})}}{\cos{({\varphi}^{(j)}_{DC})}}{\varphi}^{(j)}_{AC}(t)\right],
\label{EqA29}
\end{eqnarray}
where $\bar{E}^{(j)}_{J_s}=E_{J_s}\cos{{\varphi}^{(j)}_{DC}}$. By replacing Eq.~(\ref{EqA29}) in the Hamiltonian of Eq. (\ref{EqA20}), we obtain
\begin{eqnarray}\nonumber
\hat{\mathcal{H}} = && \frac{\omega_{q_1}}{2}\sigma_1^z+\frac{\omega_{q_2}}{2}\sigma_2^z+\omega_{r_1}a_1^\dagger a_1+\omega_{r_2}a_2^\dagger a_2-i\left[g^{(1)}_{0}+g^{(1)}_{1}\varphi^{(1)}_{AC}(t)\right]\sigma_1^y(a_1-a_1^\dagger)-i\left[g^{(2)}_{0}+g^{(2)}_{1}\varphi^{(2)}_{AC}(t)\right]\sigma_1^y(a_2-a_2^\dagger)\\
&& - i\left[g^{(3)}_{0}+g^{(3)}_{1}\varphi^{(3)}_{AC}(t)\right]\sigma_2^y(a_1-a_1^\dagger)-i\left[g^{(4)}_{0}+g^{(4)}_{1}\varphi^{(4)}_{AC}(t)\right]\sigma_2^y(a_2-a_2^\dagger),
\label{EqA27}
\end{eqnarray}
where the coupling strength 
\begin{eqnarray}
g^{(j)}_0&=&\frac{\Phi_0C^2_{c}E_{J_1}\sqrt{\frac{\hbar \omega_{r_{j}}\bar{L}_{r_{j}}}{2}}}{8\pi (C_{r_{j}}+2C_c)(C_{1}+2C_c)L_{r_{j}}\bar{E}^{(j)}_{J_s}},
\quad
 g^{(j)}_1=\frac{\Phi_0C^2_{c}E_{J_1}\sqrt{\frac{\hbar \omega_{r_{j}}\bar{L}_{r_{j}}}{2}}}{8\pi (C_{r_{j}}+2C_c)(C_{1}+2C_c)L_{r_{j}}\bar{E}^{(j)}_{J_s}}\frac{\sin{({\varphi}^{(j)}_{DC})}}{\cos{({\varphi}^{(j)}_{DC})}}\quad j=\{1,2\},
\nonumber \\
\end{eqnarray}
and
\begin{eqnarray}
g^{(j)}_0&=&\frac{\Phi_0C^2_{c}E_{J_2}\sqrt{\frac{\hbar \omega_{r_{j}}\bar{L}_{r_{j}}}{2}}}{8\pi (C_{r_{j}}+2C_c)(C_{1}+2C_c)L_{r_{j}}\bar{E}^{(j)}_{J_s}},\quad
g^{(j)}_1=\frac{\Phi_0C^2_{c}E_{J_2}\sqrt{\frac{\hbar \omega_{r_{j}}\bar{L}_{r_{j}}}{1}}}{8\pi (C_{r_{j}}+2C_c)(C_{1}+2C_c)L_{r_{j}}\bar{E}^{(j)}_{J_s}}\frac{\sin{({\varphi}^{(j)}_{DC})}}{\cos{({\varphi}^{(j)}_{DC})}}\quad j=\{3,4\} . \nonumber \\
\label{EqA31}
\end{eqnarray}

To visualize the dynamics of our system, we go to the interaction picture and make the rotating-wave approximation to obtain
\begin{eqnarray}\nonumber
\hat{\mathcal{H}}_I\approx && \frac{Ag^{(1)}_{1}}{4}\bigg[-\sigma^x_1a_1 (e^{-i\tilde{\varphi}^{(1)}_1}-e^{i\tilde{\varphi}^{(1)}_2})-\sigma^y_1a_1 (ie^{-i\tilde{\varphi}^{(1)}_1}+i e^{i\tilde{\varphi}^{(1)}_2})+\sigma^x_1a^\dagger_1(e^{-i\tilde{\varphi}^{(1)}_2}-e^{i\tilde{\varphi}^{(1)}_1})+\sigma^y_1a^\dagger_1(ie^{-i\tilde{\varphi}^{(1)}_2}+ie^{i\tilde{\varphi}^{(1)}_1})\bigg]
\\\nonumber
&& + \frac{Ag^{(2)}_{1}}{4}\bigg[-\sigma^x_1a_2(e^{-i\tilde{\varphi}^{(2)}_1}-e^{i\tilde{\varphi}^{(2)}_2})-\sigma^y_1a_2(ie^{-i\tilde{\varphi}^{(2)}_1}+ie^{i\tilde{\varphi}^{(2)}_2})+\sigma^x_1a^\dagger_2(e^{-i\tilde{\varphi}^{(2)}_2}-e^{i\tilde{\varphi}^{(2)}_1})+\sigma^y_1a^\dagger_2(ie^{-i\tilde{\varphi}^{(2)}_2}+ie^{i\tilde{\varphi}^{(2)}_1})\bigg]
\\\nonumber
&&+\frac{Ag^{(3)}_{1}}{4}\bigg[-\sigma^x_2a_1(e^{-i\tilde{\varphi}^{(3)}_1}-e^{i\tilde{\varphi}^{(3)}_2})-\sigma^y_2a_1(ie^{-i\tilde{\varphi}^{(3)}_1}+ie^{i\tilde{\varphi}^{(3)}_2})+\sigma^x_2a^\dagger_1(e^{-i\tilde{\varphi}^{(3)}_2}-e^{i\tilde{\varphi}^{(3)}_1})+\sigma^y_2a^\dagger_1(ie^{-i\tilde{\varphi}^{(3)}_2}+ie^{i\tilde{\varphi}^{(3)}_1})\bigg]
\\\nonumber
&&+\frac{Ag^{(4)}_{1}}{4}\bigg[-\sigma^x_2a_2(e^{-i\tilde{\varphi}^{(4)}_1}-e^{i\tilde{\varphi}^{(4)}_2})-\sigma^y_2a_2(ie^{-i\tilde{\varphi}^{(4)}_1}+ie^{i\tilde{\varphi}^{(4)}_2})+\sigma^x_2a^\dagger_2(e^{-i\tilde{\varphi}^{(4)}_2}-e^{i\tilde{\varphi}^{(4)}_1})+\sigma^y_2a^\dagger_2(ie^{-i\tilde{\varphi}^{(4)}_2}+ie^{i\tilde{\varphi}^{(4)}_1})\bigg] .
\end{eqnarray}
By considering $\tilde{\varphi}^{(j)}_1=\pi$, and $\tilde{\varphi}^{(j)}_2=2\pi$, we obtain
\begin{eqnarray}\nonumber
&&\hat{\mathcal{H}}_I\approx \frac{Ag^{(1)}_{1}}{2}\sigma^x_1(a_1+a^\dagger_1)+\frac{Ag^{(2)}_{1}}{2}\sigma^x_1(a_2+a^\dagger_2)+\frac{Ag^{(3)}_{1}}{2}\sigma^x_2(a_1+a^\dagger_1)+\frac{Ag^{(4)}_{1}}{2}\sigma^x_2(a_2+a^\dagger_2),
\end{eqnarray}
which corresponds to the two-qubit two-mode quantum Rabi model. We highlight that the coupling strengths $Ag_1^{(k)}/2$ depend on the external flux amplitude $A$, such that they can be adiabatically changed in order to perform the W-state generation protocol proposed in the main text.

\section{Lindblad master equation for numerical simulation}
We use the following Lindblad form master equation
\begin{eqnarray}\label{lind}
\dot{\rho}=&&-i[H_{pq},\rho]+\sum_{i=1}^M \frac{\kappa}{2}(2a_i\rho a_i^\dag- a_i^\dag a_i\rho-\rho a_i^\dag a_i)+\sum_{j=1}^2 \frac{\gamma_j}{2}(2\sigma_j\rho \sigma_j^\dag- \sigma_j^\dag \sigma_j\rho-\rho \sigma_j^\dag \sigma_j)\nonumber\\&&+\sum_{j=1}^2 \gamma_{j\phi}(\sigma_{jz}\rho \sigma_{jz} - \rho)
\end{eqnarray}
to carry out the numerical simulation. Here, $\kappa$ is the photon decay rate of the $i$th CWR, consisting of the intrinsic  part $\kappa_{in}$ and coupling part $\kappa_c$ with respect to the TL. $\gamma_j$  and $\gamma_{j\phi}$ are the energy relaxation rate and the dephasing rate of the $j$th SQ, respectively. Although the ultrastrong coupling regime is reached, we use this Lindblad form master equation because the maximum coupling strength $g$ we reach is just $0.29$ and $\kappa_{in}=10^{-4}\omega$, $\gamma_j=10^{-5}\omega$,  $\gamma_{j\phi}=10^{-4}\omega$, which are extremely small and when $\kappa_{c}=10^{-1}\omega$ is turned on, the qubit Bell state has been generated, being decoupled from the photon mode. We have testified it numerically by using a Markovian master equation \cite{fb,arm}
\begin{equation}
\dot{\rho}(t) = -i[H_{pq},\rho(t)] + \sum_{j,k>j} \Gamma_{m}^{jk} D(\ket{j}\bra{k}) \rho(t), \nonumber \\
\end{equation}
which corresponds to amplitude damping. Here, $\{\ket{j} \}_{j=0,1,2..}$ are the eigenvectors of Hamiltonian $H_{pq}$, with $H\ket{j}= \epsilon_{j}\ket{j}$, and ${D(O)\rho = 1/2(O \rho O^{\dagger} - O^{\dagger}O \rho - \rho O^{\dagger}O)}$. Also, index $m = \{1, ... , M, M+1, ...,  M + N \}$ runs through all resonators and qubits, such that $m \leq M$ refers to photon modes and $M < m \leq M + N$ refers to qubits. The decay rates, $\Gamma_{m}^{jk}$, are taken as
 \[   
\Gamma_{m}^{jk} = 
     \begin{cases}
       \kappa_{m} \frac{\delta_{kj}}{\omega}|C_{kj}^{m}|^{2}. \quad\  C^{m} = a_{m} + a_{m}^{\dagger} & \textrm{for} \quad  \quad\ m \leq M\\
       \gamma_{m} \frac{\delta_{kj}}{\omega}|C_{kj}^{m}|^{2}.  \quad\  C^{m} = \sigma_{mx}& \textrm{for} \quad \quad\ M < m \leq M + N \\
     \end{cases}
\]
 where $\Delta_{kj} = \epsilon_{k} - \epsilon_{j}$ and $C_{kj}^{m} = \langle k|C^m|j\rangle$. $\kappa_m$ is the photon decay rate of the $m$-th CWR, while $\gamma_m$  is the energy relaxation rate of the $m$-th qubit. The numerical results is almost the same as that obtained form the corresponding Lindblad form master equation Eq. \eqref{lind}.

\end{widetext}

\end{document}